\documentclass[final,3p,times,twocolumn]{elsarticle}

\usepackage{graphicx}

\usepackage{rotating}

\usepackage{lineno}

\journal{Nuclear Instruments and Methods in Physics Research Section A}

\begin{document}

\begin{frontmatter}



\title{An einzel lens with a diagonal-slit central electrode to combine steering and focusing of a low energy
ion beam}

\author[label1]{P.~Mandal \corref{cor1}}
\ead{drupm@iacs.res.in}
\author[label2]{G.~Sikler \corref{cor2}}
\author[label1]{M.~Mukherjee}
\address[label1]{Raman Center for Atomic Molecular and Optical Sciences,
Indian Association for the Cultivation of Science, 2A \& 2B Raja
S.~C.~Mulllick Road, Kolkata 700032, India}

\address[label2]{Gesellschaft f\"{u}r Schwerionenforschung, Planckstr. 1, 64291 Darmstadt, Germany}
\cortext[cor1]{Corresponding author}
\cortext[cor2]{Present address: Babcock Noell GmbH Alfred-Nobel-Strasse 20, 97080 Wuerzburg, Germany}

\begin{abstract}
In many applications of the simple three-element einzel lens, such
as injecting a low energy ion beam into a high-field Penning trap,
there is a need for small-angle steering as well as focusing of the
beam. We have analyzed a diagonal-slit cylinder serving as the
middle electrode of such a lens and have shown that such an
electrode configuration significantly diminishes the aberration
associated with such a deflection.
\end{abstract}


\end{frontmatter}

\section{Introduction}
\label{sec1} The einzel lens, consisting of a single activated
cylindrical electrode centered between two cylindrical electrodes at
ground~\cite{ada72}, is a particularly useful electrostatic lens for
low energy ion beams. This is because it is simple, compact and
azimuthally symmetric. It is particularly useful for applications
such as the injection of a low energy ion beam into a high magnetic
field Penning trap, where a compact system is needed for focusing a
low energy ion beam onto the azimuthally symmetric magnetic field
lines of the trap~\cite{muk08}.

However, it is difficult, in general, to produce the azimuthally
symmetric focussing quadrupole field in an einzel lens without
strong components of higher order multipoles. Most significant of
these is the azimuthally symmetric octupole which results in
spherical aberration. Many studies of this aberration have been
carried out, one of the most thorough being Ref.~\cite{rem85}. As a
result of this aberration only about a quarter of the internal
diameter of the central electrode of a simple einzel lens is useful
if the emittance of a typical low-energy ion beam is not to be
significantly increased. This means that an einzel lens has to be
considerably larger than the beam it has to focus, making it less
compact than desired.

This problem is compounded when the beam has to be steered as well
as focused, as is necessary for injection into a Penning trap. This
is because any misalignment of the ion trajectories with the
magnetic field lines of the trap results in cyclotron motion that
becomes amplified as the ions enter the high field of the trap,
interfering with the use of the Penning trap as a mass measuring
device. The enlargement of the einzel lens to accommodate its
spherical aberration leaves little room for separate electrostatic
steering electrodes. It is therefore desirable to combine the
steering and focusing functions in the single central electrode of
the einzel lens.

However, the introduction of a steering field by simply segmenting
the central electrode of an einzel lens can also bring with it an
aberrant field that can itself distort the emittance diagram at the
beam focus. This paper describes a means of reducing that aberrant
field.

\section{The aberration of a central steering field of an einzel lens}
\label{sec2}

Parallel plate electrodes replacing the middle cylinder in an einzel
lens, may be used to combine the deflecting field with the focusing
filed. Since there is no azimuthal symmetry in the focusing field,
such a design is useless for practical purposes. However, the design
can be modified by placing another set of parallel plates, thus
forming a box at the middle. Such a configuration, to a close
approximation, produces an azimuthally symmetric quadrupolar
potential required for focussing. Each of the four plates may be
activated independently with different potentials in order to
achieve bending in any direction. However, the deflecting field of a
box electrode would be highly non-linear, being much stronger in the
direction of steering on the axis of the lens than near the surfaces
of the electrodes that are parallel to the steering direction. Thus
the design of such a system is not desirable when an ion beam of a
given emittance is required to be steered by sufficiently larger
angle while being tightly focused.

Another possible way to introduce the deflecting field with the
focusing field is to make a parallel slit in the middle cylinder in
an einzel lens as it is shown in fig.~1a. The deflecting field in
this design will increase along the slit axis perpendicular to the
direction of propagation resulting in significant sextupole
distortion. The design may be improved to minimize the distortion by
slitting the central cylinder into four segments as shown in
fig.~1b. It thus suggests that there should be $8-$sector
segmentations if it is required to steer the ion beam in any
arbitrary direction and the design of such a system will be
complicated.

\begin{center}
\begin{figure}
\label{fig1}
\vspace{-0.5cm}
\includegraphics[width=0.4\textwidth]{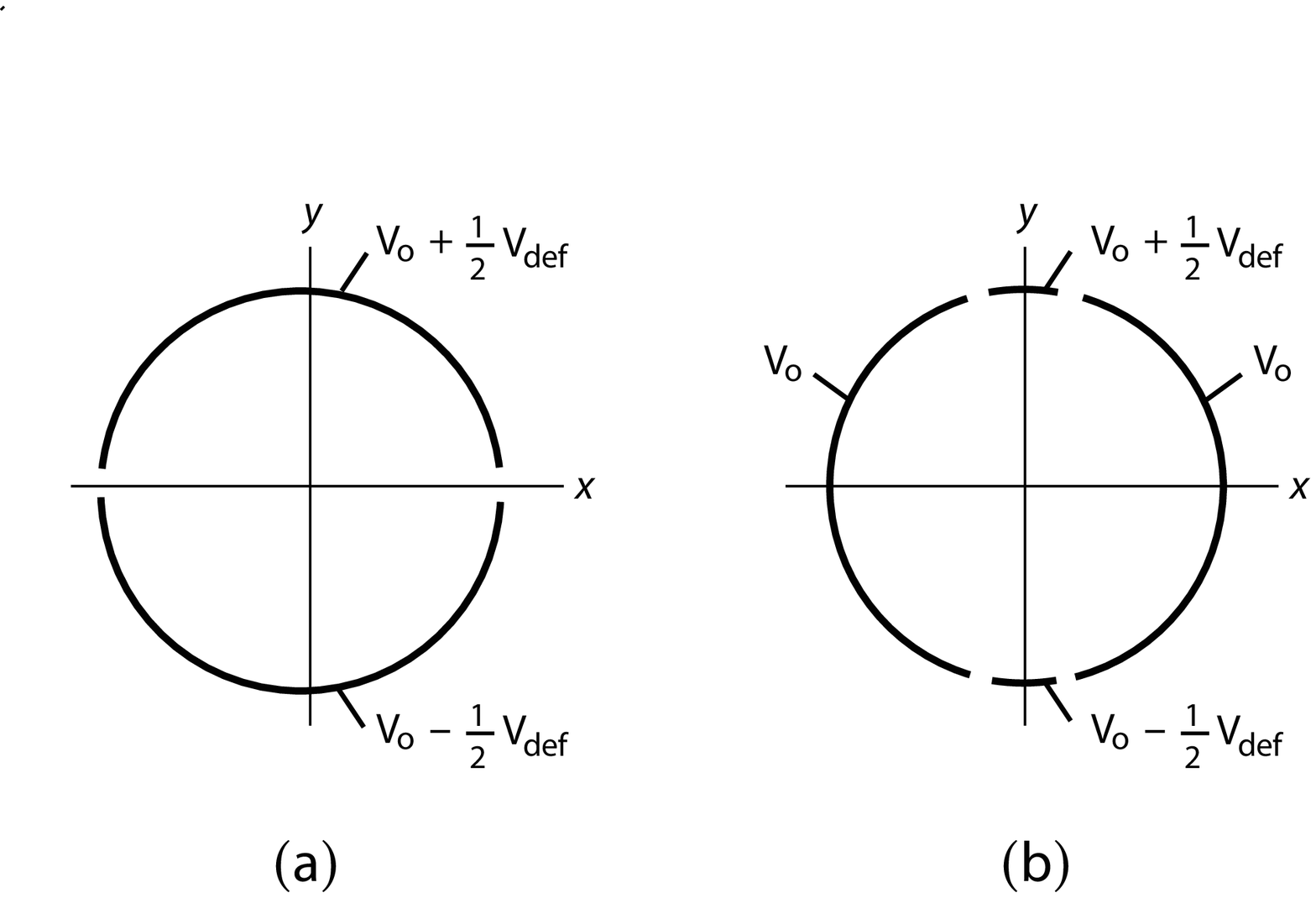}
\caption {Two different segmentations of the central electrode of an
einzel lens for steering. $V_{0}$ is the potential of the central
electrode required for focusing. (For the same deflection the
deflecting potential, $V_{def}$, for the two cases will be
different.}
\end{figure}
\end{center}

\begin{center}
\begin{figure}
\label{fig2}
\vspace{0.5cm}
\hspace{0.5cm}
\includegraphics[width=0.4\textwidth]{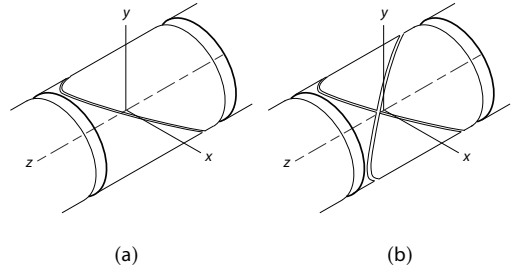}
\caption {Diagonal segmentation of the central electrode of an
einzel lens for steering, (a) for the $Y$ direction only and (b) for
both $X$ and $Y$ directions.}
\end{figure}
\end{center}
\vspace{-1.5cm}
Here we show that an einzel lens with the middle cylinder having a
diagonal slit as shown in fig.~\ref{fig2}a, is the best solution to
this problem. The design allows comparatively larger steering and
focussing of an ion beam within acceptable aberration. The cylinder
may have another diagonal slit perpendicular the first one (as in
fig.~\ref{fig2}b) so as to achieve the steering along any direction
on the focal plane. Ref.~\cite{sik03} introduces such a
configuration on its first implementation at the
SHIPTRAP~\cite{blo07} facility at GSI and it is in use also at TITAN
facility TRIUMF~\cite{dal07}. Here we have studied the design thoroughly and
compared it with a parallel-slit configuration. At the end we show
that a diagonal-slit is the best choice compared to other design in
view of the maximum extent of the deflection angle that it allows to
focus a low energy ion beam without significant aberration.

\section{Simulation results and discussions}
\label{sec3}

\begin{center}
\begin{figure}
\label{fig3} \hspace{-1cm}
\includegraphics[width=0.5\textwidth, angle=270]{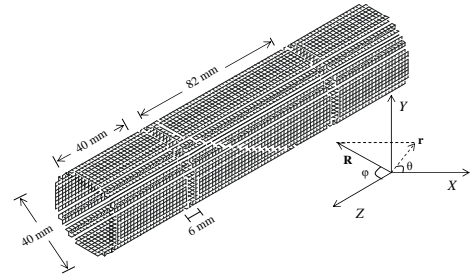}
\vspace{-4cm}
\caption {A schematic of a diagonal-slit deflector and the
coordinate system used in the simulation. End cylinders are
identical.}
\end{figure}
\end{center}

In order to evaluate the deflection aberration of a diagonal cut
einzel lens we have taken a typical lens geometry shown in
fig.~\ref{fig3}. This einzel lens has a middle electrode of length
$82$~mm, inner diameter $40$~mm and two identical electrodes at each
end of length and inner diameter $40$~mm. The middle cylinder is
separated by a distance $6$~mm from each end cylinder. Beam focusing
is achieved by elevating the potential of the middle electrode (for
positive ions) while keeping the end electrodes at ground.

\vspace{-1cm}

A representative beam has been taken to study the performance of
different ion optical systems. Each ion has a mass $100$~amu,
positive charge of one electron unit and a kinetic energy of
$100$~eV. The array of ions start at $137$~mm from the center of the
lens ($50$~mm before the entrance to the first ground electrode).
They are initially spaced symmetrically about $y=0$ within $10$~mm
along $Y$ axis, in steps of $1$~mm with their divergence ($y'$)
symmetrically spaced about $y'=0$ within $4.4$~mrad, in steps of
$0.44$~mrad. Thus the beam initially forms a square array of $11$ by
$11$ points with $121$ ions in the $y - y'$ emittance diagram having
an emittance area of $44$~mm~mrad. Each ion in this diagram has
neither $x-$ displacement nor $x-$divergence distribution
(\emph{i.e.} $x=0$~mm, $x'=0$). The trajectory of the ion beam has
been simulated using SIMION $3$D~$7.0$, including the grounded
vacuum enclosure of the lens assembly. The potential at the middle
electrode is then varied until the array is focused at a desired
distance from the lens assembly. In our simulation the focus is set
at an axial distance of $233$~mm from the lens center ($146$~mm from
the lens exit) when the central electrode in fig.~\ref{fig3} is
elevated at potential of $52$~V. The emittance diagrams at the focus
are examined for aberrations and the size of the emittance arrays
adjusted until the spherical aberration is regarded as acceptable.
This is found to be the case with the emittance array described
earlier. The resulting $y-y'$ emittance diagram at the focus is
shown in fig.~4. This clearly shows the ``S" shaped distortion due
to the octupole aberration but that is regarded as acceptable since
it doesn't significantly increase the effective area of the
emittance diagram.

\begin{center}
\begin{figure}
\vspace{-0.5cm}
\includegraphics[width=0.5\textwidth]{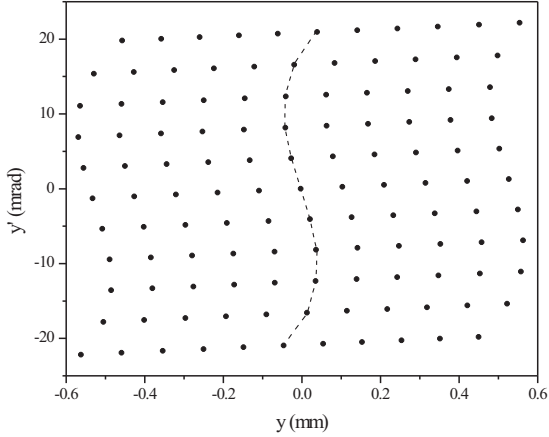}\label{fig4}
\vspace{-1cm}
\caption {The $y-y'$ emittance diagram at the focus of
an einzel lens for potential of $52$~V. The dotted line (only to
guide eye) clearly shows an ``S" shaped distortion symmetric about
$y=0$, $y'=0$.}
\end{figure}
\end{center}
\vspace{-1cm} Because of the azimuthal symmetry of an einzel lens
the $x -x'$ emittance diagram at the focus is, of course, identical
to fig.~4.

Now with the same initial beam emittance, we have studied the effect
of combining the deflecting field with the focusing field. The
middle cylinder in the einzel lens is slit in parallel along the
$XZ$ plane into two halves and each of them is activated
independently with different voltages so as to achieve the steering
along $Y$ direction. A potential difference of $4.1$~V about $52$~V
focusing potential is applied between the middle electrodes in order
to obtain a steering angle of $86$~mrad of the same representative
beam. The $y-y'$ emittance diagram at the focus is shown in fig.~5
which shows a large ``U" shaped distortion representing a clear
signature of an unacceptable sextupole aberration.
\begin{center}
\begin{figure}[hhh]
\label{fig5} \vspace{-0.5cm}
\includegraphics[width=0.5\textwidth]{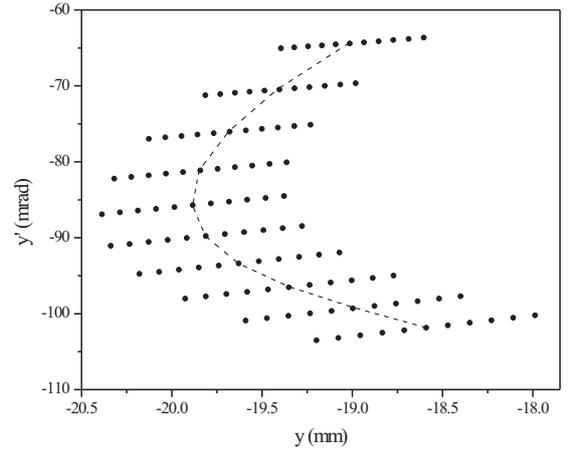}
\vspace{-1cm} \caption {The $y-y'$ emittance diagram at the focus of
a parallel-slit deflector with deflecting potential $4.1$~V. The
dotted line showing a clear ``U" shape, connects the points that
start with initial $y=-5$ mm to $y=5$ mm and $y'=0$.}
\end{figure}
\end{center}
\vspace{-1cm}

The central cylinder of the einzel lens is segmented by a diagonal
cut as in fig.~\ref{fig3}. The potential difference between the two
segments is adjusted so as to achieve the same bending of $86$~mrad
as for the parallel-slit electrode. This is achieved with a
potential difference of $6$~V (as compared to the $4.1~$V required
for the parallel segmented electrode, due to the smaller dipole
field of the diagonally segmented electrode.) The resulting $y - y'$
emittance diagram at the focus is shown in fig.~6a. The signature of
the deflection field aberration is barely detectable in this figure.
In order to make it prominent, another simulation has been performed
for a deflection potential of $10$~V resulting in a steering by
$140$~mrad. The $y-y'$ emittance diagram is presented in fig.~6b.
Fig.~6a and fig.~6b thus represent a qualitative comparison of the
sextupole distortion in the emittance diagram and hence the
aberration as a function of the deflecting potential or the steering
angle.

\begin{center}
\begin{figure*}[ttt]
\label{fig6} \vspace{-0.5cm}
\includegraphics[width=\textwidth]{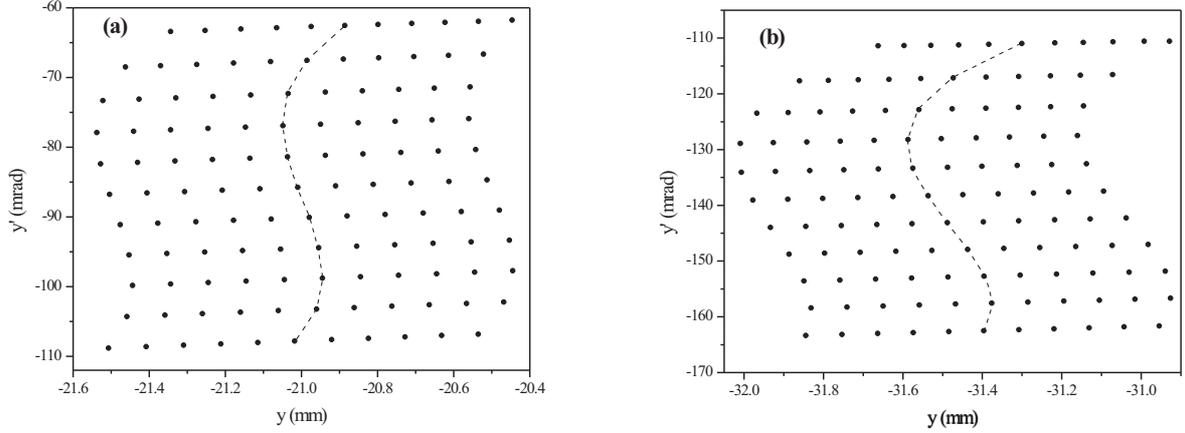}
\vspace{-1cm} \caption {The $y-y'$ emittance diagram at the focus of
a diagonal-slit deflector for a deflecting potential of (a) $6$ V
and (b) $10$ V. The dotted lines (only to guide the eye) connecting
the points having initial $y-$distribution and $y'=0$ show nearly
``S" shaped distortion. These lines are not rotationally symmetric
about the center representing thus higher order distortions.}
\end{figure*}
\end{center}
\vspace{-1cm}

Fig.~5 and fig.~6a present a comparative performance between a
parallel-slit and a diagonal-slit configurations of same dimensions
for a given angle of steering of a low energy ion beam. For
diagonal-slit deflector, the ions are converged within $1$~mm
(fig.~6a) while for the parallel-slit deflector they spread over a
region of $2.4$~mm (fig.~5) along the axis of steering. These two
figures thus naturally inspire to choose a diagonal-slit
configuration as the best compact low energy ion beam
deflector-lens.

In order to realize the reason behind the better performance of a
diagonal-slit rather than a parallel-slit configuration, the
azimuthal symmetry associated with these two configurations should
be considered. The distortion for the deflecting field in the middle
electrodes changes in the axial direction for a diagonal-slit
deflector while it is constant for a parallel-slit design. At the
entrance and exit of the diagonal-slit deflector, the deflecting
field has a curvature such that the field is diminished with the
distance along the axis in the direction perpendicular to the field.
The field curvatures are in opposite directions relative to the
center of the middle electrodes and therefore there occurs a
cancelation of the distortion. The cancelation, however will not be
perfect since the opposite deviations occur in separate regions of
the beam and may not be equal. They only tend to cancel each other
because the beam does not change much as it passes through the
central electrode, the degree of cancelation being determined by the
length to diameter ratio of the middle electrode and the initial
transverse emittance of the ion beam. At the center of the electrode
a strong deflecting field exists along which direction the beam is
to be steered. The solution of Laplace's equation inside the
deflector-lens system can be represented as
\begin{equation}
\label{eqn1}
V=\sum_{l,m}{a_{lm}e^{im\phi}R^{l}P^{m}_{l}(\cos\theta)},
\end{equation}
where $a_{lm}$ indicates the strength of $l, m$ multipole and
$P^{m}_{l}(\cos\theta)$ is the Legendre polynomial of $\cos\theta$.
The ions are focused due to the azimuthally symmetric quadrupole
component ($l=2$, $m=0$) of this potential in case of an einzel
lens. They are deflected due to the dipole component ($l=1$, $m=1$)
for a deflector. Several higher order multipoles contribute
significantly in any practical ion-optic system resulting in
different order of aberrations. The first order aberration arises
from the azimuthally symmetric octupole part ($l=4$, $m=0$) in an
einzel lens and the next higher order contribution arises from the
sextupole component corresponding to $l=3$, $m=1$ for a
diagonal-slit deflector. Considering these four components of the
potential, the transverse components of the electric field at the
middle ($z=0$) of a real diagonal-slit deflector-lens (for steering
along $Y$) are given by
\begin{equation}
\label{eqn2}
E_{y}=-a_{11}+a_{20}y+\frac{3}{2}a_{31}(x^{2}+3y^{2})-\frac{3}{2}a_{40}y^{3}
\end{equation}
and
\begin{equation}
\label{eqn3}
E_{x}=a_{20}x+3a_{31}xy-\frac{3}{2}a_{40}x^{3}
\end{equation}
In presence of the electric field in Eqn.~\ref{eqn2}, the
$y-$divergence ($y'$) of an ion beam having initial $x,x'=0$
initially, is approximately represented as a function of their
instantaneous $y-$position at the center~($z=0$)~of the
deflector-lens in the following way
\begin{equation}
\label{Eqn4}
y'=a+by+cy^{2}+dy^{3},
\end{equation}
where $a$, $b$, $c$, $d$ are proportional to $a_{11}$, $a_{20}$,
$a_{31}$ and $a_{40}$ respectively. When a parallel beam ($x'=y'=0$
initially) of ions having only initial $y-$ distribution symmetric
about $y=0$, is allowed to pass through the diagonal-slit deflector,
the emittance of the beam follows Eqn.~\ref{Eqn4}. The $y-y'$
emittance of the beam is observed at the center of the diagonal-slit
deflector-lens for a deflecting potential $10$~V and it is shown in
fig.~7. The plot is fitted with the Eqn.~\ref{Eqn4} with $a=-100.977
(011)$, $b=0.198 (004)$, $c=-0.04 (001)$, $d=-0.005 (0001)$. It
shows that the magnitude of the coefficients of successive higher
order multipoles gradually decreases but their contributions are
more dominating near the surface of the electrodes.
\begin{center}
\begin{figure}
\label{fig7} \vspace{-0.4cm}
\includegraphics[width=0.5\textwidth]{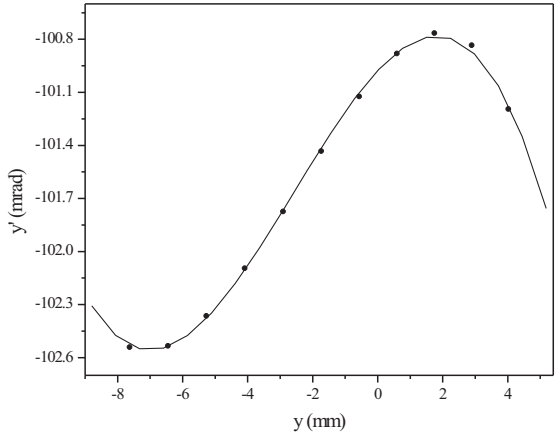}
\vspace{-1cm} \caption {The $y-y'$ emittance diagram at the center
of the diagonal-slit deflector with deflecting potential $10$~V. The
plot is fitted to a third order polynomial as explained in the
text.}
\end{figure}
\end{center}
\vspace{-1cm}

\begin{center}
\begin{figure}
\label{fig8} \vspace{-0.4cm}
\includegraphics[width=0.5\textwidth]{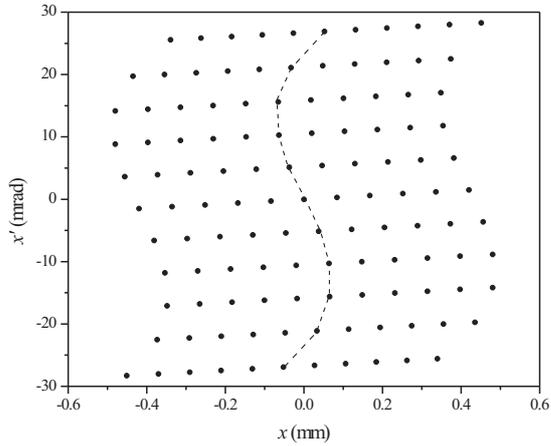}
\vspace{-1cm} \caption {The $x-x'$ emittance diagram at the focus of
a diagonal-slit deflector with deflecting potential $10$~V. Dotted
line shows an ``S" shaped distortion symmetric about the center.}
\end{figure}
\end{center}
\vspace{-1cm}
However, as the beam propagates through the system and
is converged, the emittance plot at the focus takes the shape as it
is shown by dotted line in fig.~6b. In case of an einzel lens the
first and third terms in Eqn.~\ref{Eqn4} are absent and hence the
emittance diagram shows only an ``S" shaped distortion at the focus
as represented in fig.~4. Since the coefficient $a_{31}$ is
proportional to the deflecting potential, the distortion due to the
sextupole component increases almost linearly with increasing
deflecting potential.

Eqn.~\ref{eqn3} predicts that a representative beam having only
$x-$emittance and $y=y'=0$ initially, experiences a distortion only
due to the octupole component and the deflecting potential has no
effect in the emittance diagram. This point has also been checked in
our simulation. Fig.~8 shows the $x-$emittance diagram at the focus
of the diagonal-slit deflector-lens with deflecting potential $10$~V
for this representative beam. It does not show any significant
difference as compared to fig.~4. A small deviation is the
consequence of the contribution of higher order multipoles of little
interest and beyond the scope of this study.

\section{Conclusion}
A novel system has been described for using the central electrode of
a simple einzel lens for both steering and focusing a low energy ion
beam with minimal aberrations. It involves cutting the central
electrode in planes that  pass through the lens center and are
orthogonal to each other but that are tilted to the lens axis. The
position of the focus and the steering angle in any direction can
then both be controlled by adjusting potentials on the middle
electrodes. The system is very useful to handle a steering of the
order of $10$ degrees or less. Furthermore, since $x$ and $y$
coordinates are independent, two diagonal cuts at $90$ degrees
azimuth to each other (fig.~\ref{fig2}b), can be used to steer the
beam in any arbitrary direction.

\textbf{Acknowledgments} The authors sincerely thank Prof.
R.~B.~Moore, McGill University, Montreal, Canada for his useful
suggestions at different stages of this work and for his assistance
in drafting of this manuscript. P.~Mandal is thankful to the Council
of Scientific and Industrial Research (CSIR), India for sponsoring
the fellowship during the research work.

\end{document}